# Higgs Phenomenology in Supersymmetric $SU(3)_C \otimes SU(3)_L \otimes U(1)_X$ model

Sutapa Sen and Aparna Dixit

Department of Physics, Christ Church P.G. College, Kanpur 208001,U.P.India

## ABSTRACT

We build a supersymmetric (SUSY) version of a recently proposed $SU(3)_C \otimes SU(3)_L \otimes U(1)_X$ model with heavy charged lepton and no bilepton gauge bosons. The model is an anomaly- free three- generation extension of the Standard Model with $SU(3)_L \otimes U(1)$ gauge symmetry. The scalar sector of the 3-3-1 model is analyzed to obtain mass spectra for neutral and charged Higgs bosons and their interactions with gauge bosons, quarks, leptons. The tree-level decay widths of Higgs bosons at TeV energies are presented.

## 1. Introduction

New physics at TeV scale can be probed by upcoming experiments at Fermilab Tevatron and CERN Large Hadron Collider(LHC) which offer crucial tests for the minimal supersymmetric extension of the standard model (MSSM)[1]. The scalar sector of the model remains without experimental confirmation of the elusive Higgs boson [2] and there exist the possibilities of realizations of mechanism of spontaneous electroweak symmetry breaking and hierarchy mass problem. A simple extension of SM gauge symmetry at TeV energies is $SU(3)_C \otimes SU(3)_L \otimes U(1)_X$ ( 3-3-1). While several versions of 3-3-1 models exist in literature[3,4], these can be characterized by the embedding of electric charge operator in $SU(3)_L$ generators. The 3-3-1 models can be embedded in 3-4-1 gauge symmetry [5] with the most general charge operator

$$\frac{Q}{e} = T_{3L} + aT_{8L} + bT_{15L} + XI_4 \qquad (1)$$

where $T_{iL} = \frac{\lambda_{iL}}{2}$, $i = 3, 8, 15$; $\lambda_{iL}$ are Gell-Mann matrices for $SU(4)_L$ group.

The supersymmetric extension has been considered for 3-3-1 models [6] with extended Higgs sector. The scalar sector differs from MSSM and includes new trilinear self-couplings for Higgs bosons to provide SUSY breaking. The phenomenological consequences predicted are distinct from MSSM for gauge boson and scalar sectors for these 3-3-1 models.

The main motivation of the present work is to build the SUSY version of a 3-3-1 model [7] which provides a simple extension of MSSM at TeV energies. We consider a three-generation 3-3-1 model [8] without bilepton gauge bosons derived from 3-3-1-1 gauge symmetry which is a subgroup of $SU(4)_{PS} \otimes SU(4)_{L+R}$. The electric charge operator is

$$\frac{Q}{e} = T_3 - \sqrt{3} T_8 + \sqrt{6} T_{15} + \frac{(B-L)}{2} I_4 \qquad (2)$$

where $T_i$, $i = 3, 8, 15$ are diagonal generators of $SU(4)_{L+R}$ and $I_4$ is the 4 x 4 identity matrix.

The coupling constants of $SU(3)_L$ and $U(1)_X$, $g_L$ and $g_X$ are related to electroweak mixing angle as $g_X^2/g_L^2 = \sin^2\theta_W / (1 - 4\sin^2\theta_W)$ [9]. The SUSY version of the model has an extended scalar sector with three Higgs triplets along with three anti-multiplets to cancel chiral anomalies generated by Higgsinos. There are two important points of differences from the other SUSY 3-3-1 models [6]

- There are no bilepton gauge bosons since (B-L) number is conserved and (B-L) is an operator of the $SU(4)_{PS}$ group.
- The trilinear Higgs scalar couplings which are a special feature of the other versions of the 3-3-1 models are absent in the present case.

The non-supersymmetric 3-3-1 model with heavy charged lepton [8] is reviewed in Sec2

In Sec.3 we present the superfield content of the model, along with the superpotential. Sec.4 deals with the scalar potential and constraint equations. Sec.5 deals with the mass spectrums of the six CP-even and CP-odd neutral scalars, single and double- charged scalars. Sec.6 deals with the interactions of Higgs scalars with gauge bosons and fermions. The decays H→WW, H→f $\bar{f}$ are considered at tree level. Finally Sec.7 is a short discussion on results and conclusions.

## 2. The non-supersymmetric 3-3-1 model

We consider the 3-3-1 gauge group to be embedded in $SU(4)_{PS} \otimes SU(4)_{L+R}$ which breaks to $SU(3)_C \otimes SU(3)_L \otimes U(1)_{B-L} \otimes U(1)_{Y1}$. Further breaking of $U(1)_{B-L} \otimes U(1)_{Y1} \rightarrow U(1)_X$ where $X = Y_1 + (B-L)/2$ leads to 3-3-1 model with exotic charged particles. We consider 3-3-1-1 gauge symmetry $G_C \otimes G_L$ with five basic fundamentals:

$G_C = SU(3)_C \otimes U(1)_{B-L}$: $M = (3_C, 1/6)$; $N = (1_C, -1/2)$

$G_L = SU(3)_L \otimes U(1)_{Y1}$: $a = (3_L, 1/2)$; $b = (1_R, -3/2)$; $c = (1_R, -1/2)$ (3)

The matter multiplets include bifundamental structures for scalars and fermions

*Quarks* : $Q_i = (d_i, u_i, D_i)^T_L \sim (3_C, 3_L^*, -1/3) = a*M$; $i = 1,2$; $Q_3 = (t, b, T)^T_L \sim (3_C, 3_L, 2/3) = aM$

$u^C_{Rj} \sim (3_C^*, 1_R, -2/3) = cM^*$; $d^C_{Rj} \sim (3_C^*, 1_R, 1/3) = c^*M^*$; $j = 1,2,3$

$D^C_{Ri} \sim (3_C^*, 1_R, 4/3) = b^*M^*$, $i = 1,2$; $T^C_R \sim (3_C^*, 1_R, -5/3) = bM^*$;  (4)

*Leptons*: $L_\alpha = (\nu_\alpha, l_\alpha, P_\alpha^+) \sim (1_C, 3_L, 0) = aN$; $\alpha = e, \mu, \tau$.

*Singlets*: $e_{R\alpha}^C \sim (1_C, 1_R, 1) = c^*N^*$; $P_{R\alpha}^C \sim (1_C, 1_R, -1) = bN^*$; $\alpha = e, \mu, \tau$  (5)

*Scalars*: $\eta = (\eta^0, \eta_1^-, \eta_2^+)^T \sim (1_C, 3_L, 0) = ac$; $\rho = (\rho^+, \rho^0, \rho^{++})^T \sim (1_C, 3_L, 1) = ac^*$

$\chi = (\chi^-, \chi^{--}, \chi^0)^T \sim (1_C, 3_L, -1) = ab$  (6)

For supersymmetric 3-3-1 models [6], the scalar sector is extended by adding extra Higgs triplets in respective anti-multiplets to cancell chiral anomalies generated by the Higgsinos

$\eta' = (\eta^{0'}, \eta_1^{+'}, \eta_2^{-'})^T \sim (1_C, 3^*_L, 0) = a^*c^*$; $\rho' = (\rho'^-, \rho'^0, \rho'^{--})^T \sim (1_C, 3^*_L, -1) = a^*c$

$\chi' = (\chi'^+, \chi'^{++}, \chi'^0)^T \sim (1_C, 3^*_L, 1) = a^*b^*$  (7)

The vacuum expectation values [VEV] are denoted by $<\eta^0> = v/\sqrt{2}$; $<\eta'^0> = v'/\sqrt{2}$;

$<\rho^0> = u/\sqrt{2}$; $<\rho'^0> = u'/\sqrt{2}$; $<\chi^0> = V/\sqrt{2}$; $<\chi'^0> = V'/\sqrt{2}$  (8)

We now introduce chiral superfields $\hat{\varphi}$ by extending the particle content to include squarks, sleptons and Higgsinos where $\varphi = Q_{1,2,3}, L_{e,\mu,\tau}, u_R^C{}_{1,2,3}, d_R^C{}_{1,2,3}, D_R^C{}_{1,2}, T_R^C, e_R^C{}_{1,2,3}, P_R^C{}_{1,2,3}, \eta, \eta',$ $\rho, \rho', \chi, \chi'$ i.e., 27 chiral superfields The vector superfields for the gauge bosons of each of $SU(3)_C$, $SU(3)_L$ and $U(1)_X$ are [6] $\hat{V}_C, \overline{\hat{V}}_C, \hat{V}, \overline{\hat{V}}$ and $V_X$ where $\hat{V}_C = T^i \hat{V}^i{}_C$, $\hat{V} = T^i \hat{V}^i$, $T^i = \lambda_i/2$; $\overline{\hat{V}}_C = \overline{T}^i \hat{V}^i{}_C, \overline{\hat{V}} = \overline{T}^i \hat{V}^i$, $\overline{T}^i = -\lambda_i^*/2$. Here $\lambda_i$ denotes Gell-Mann matrices, $i = 1,2,\ldots 8$.

## 3. Supersymmetric 3-3-1 model

The Lagrangian of the SUSY 3-3-1 is of two parts, the SUSY generalization of 3-3-1[6] and the SUSY breaking term

$$L_{3-3-1} = L_{SUSY} + L_{breaking} \quad (9)$$

$$L_{SUSY} = L_{lepton} + L_{quark} + L_{gauge} + L_{scalar} \tag{10}$$

The $L_{quark}$ and $L_{gauge}$ terms are identical to that of Ref.[6]. The lepton and scalar terms differ in particle content and superpotential from [6]. The superpotential W of the model consistent with gauge symmetries is given by

$$W = \frac{W_2}{2} + \frac{W_3}{3} \tag{11}$$

The bilinear terms are

$$W_2 = \mu_\eta \hat{\eta} \hat{\eta}' + \mu_\rho \hat{\rho}\hat{\rho}' + \mu_\chi \hat{\chi}\hat{\chi}' \tag{12}$$

The trilinear terms include

$$W_3 \text{ (trilinear)} = \Sigma_a k^e_a \hat{L}_a \hat{\rho}' \hat{e}_{Ra}{}^c + \Sigma_a k^P_a \hat{L}_a \hat{\chi}' \hat{P}_{Ra}{}^c + \sum_{i\alpha}[k^d_{i\alpha} \hat{Q}_i \hat{\eta} \hat{d}_{R\alpha}{}^c$$
$$+ k^u_{i\alpha} \hat{Q}_i \hat{\rho} \hat{u}_{R\alpha}{}^c] + \hat{Q}_3 \sum_\alpha [k^b_\alpha \hat{\rho}' \hat{d}_{R\alpha}{}^c + k^t_\alpha \hat{\eta}' \hat{u}_{R\alpha}{}^c]$$
$$+ k^T \hat{Q}_3 \hat{\chi}' \hat{T}_R{}^c + \sum_{i\beta} (k^D{}_{i\beta} \hat{Q}_i \hat{\chi} \hat{D}_{R\beta}{}^c) \tag{13}$$

The indices $a = 1,2,3$; $i = 1,2$; $\alpha = 1,2,3$ and $\beta = 1,2$ refer to three generations of leptons and quarks. The following mass matrices are obtained from the superpotential.

$$M_u = \frac{1}{\sqrt{2}} \begin{pmatrix} k^u_{11} u & k^u_{12} u & k^u_{13} u \\ k^u_{21} u & k^u_{22} u & k^u_{23} u \\ k^t_1 v' & k^t_2 v' & k^t_3 v' \end{pmatrix} \quad M_d = \frac{1}{\sqrt{2}} \begin{pmatrix} k^d_{11} v & k^d_{12} v & k^d_{13} v \\ k^d_{21} v & k^d_{22} v & k^d_{23} v \\ k^b_1 u' & k^b_2 u' & k^b_3 u' \end{pmatrix} \tag{14}$$

$$M_D = \frac{1}{\sqrt{2}} V \begin{pmatrix} k^D_{11} & k^D_{12} \\ k^D_{21} & k^D_{22} \end{pmatrix} \quad M_T = k^T \frac{V'}{\sqrt{2}} \; ; \; M^l_{ab} = k^e_{ab} \frac{u'}{\sqrt{2}} \; ; \; M^P_{ab} = k^P_{ab} \frac{V'}{\sqrt{2}}. \tag{15}$$

The scalar sector includes six Higgses in eqn[6,7], which at $SU(2)_L$ level consists of two $SU(2)_L$ doublets $(\eta,\eta'),(\rho,\rho')$ corresponding to $(\eta,\eta') = (H_d, H_t')$; $(\rho',\rho) = (H_b',H_u)$. The lepton masses are proportional to $<\rho'> = \frac{u'}{\sqrt{2}}$. The scalars $(\eta,\eta'),(\rho,\rho')$ belong to $\underline{3}_L$, $\overline{\underline{3}}_L$ of $SU(3)_L$.

The soft terms which breaks SUSY and are consistent with (3-3-1) gauge symmetry include bilinear terms $B\mu H_1 H_2'$ instead of trilinear Higgs couplings.

$$-L_{breaking} = \Sigma_{\phi n} [m_{\phi n}{}^2 \phi_n{}^\dagger \phi_n + m_{\phi' n}{}^2 \phi_n'{}^\dagger \phi_n' + B\mu_{\phi n} \phi_n \varphi_n']$$
$$+ \Sigma_f m_{\Psi f}{}^2 \Psi_f{}^\dagger \Psi_f + \frac{1}{2} [\Sigma_\delta (m_c \lambda_c{}^\delta \lambda_c{}^\delta + m_f \lambda^\delta \lambda^\delta) + m_X \lambda_X \lambda_X + H.c]$$
$$+ \Sigma_{ab} K^e{}_{ab} \tilde{L}_a \rho' \tilde{e}_{Rb}{}^c + \Sigma_{ab} K^P{}_{ab} \tilde{L}_a \chi' \tilde{P}_{Rb}{}^c + \sum_{i\alpha} [K^d{}_{i\alpha} \tilde{Q}_i \eta \tilde{d}_{R\alpha}{}^c$$

$$+ K^u{}_{i\alpha}\tilde{Q}_i\,\rho\,\tilde{u}_{R\alpha}{}^c] + \tilde{Q}_3\sum_\alpha[K^b{}_{\alpha}\rho'\,\tilde{d}_{R\alpha}{}^c + K^t{}_{\alpha}\eta'\,\tilde{u}_{R\alpha}{}^c]$$

$$+ K^T\tilde{Q}_3\,\chi'\tilde{T}_R{}^c + \sum_{i\beta}(K^D{}_{i\beta}\,\tilde{Q}_i\chi\,\tilde{D}_{R\beta}{}^c) \tag{16}$$

The summations extend over scalars $\phi_n$ ($\phi_n = \eta, \rho, \chi$) and sfermions $\psi_f$, $f = \tilde{L}, \tilde{Q}$

The gauginos are summed over $SU(3)_C$, $SU(3)_L$ indices $\delta = 1,2..8$ while $a, b = e, \mu, \tau$

denote three generations of sleptons. The squarks are summed over $i = 1,2$; $\alpha = 1,2,3$ and

$\beta = 1,2$ generations.

The main points of differences of the present model from other 3-3-1 models are

- inclusion of $B\mu_{\phi n}\,\phi_n\,\phi_n'$, ($\phi_n = \eta,\rho,\chi$) terms in $L_{breaking}$ soft terms.

- R-parity is conserved in the model.

### 4. The Scalar Potential

The scalar potential of the model (involving $\eta, \eta', \rho, \rho', \chi, \chi'$ fields) is given by
$$V_H = V_F + V_D + V_{soft}$$
where 
$$V_F = \Sigma_a F_a{}^* F_a\,;\quad F_a = \frac{\partial W}{\partial \phi_a{}^*} \tag{17}$$

$$V_F = \Sigma_i\,\frac{|\mu_\eta\eta_i|^2 + |\mu_\eta'\eta_i'|^2 + |\mu_\rho\rho_i|^2 + |\mu_\rho'\rho_i'|^2 + |\mu_\chi\chi_i|^2 + |\mu_\chi'\chi_i'|^2}{2} \tag{18}$$

$$V_D = \frac{1}{2}(D^\alpha D^\alpha + DD),\text{ where } D^\alpha = g/2\,\varphi_i{}^\dagger \lambda_{ij}{}^\alpha \varphi_j\,;\ D = g_X\,\varphi_i{}^\dagger X \varphi_i \tag{19}$$

here $\varphi = \eta, \rho, \chi$; $\lambda^\alpha (\alpha = 1,2,..8)$ are $SU(3)_L$ generators and X denotes $U(1)_X$ charge.

$$V_D = \frac{g_X{}^2}{2}(\rho^\dagger\rho - \rho'^\dagger\rho' - \chi^\dagger\chi + \chi'^\dagger\chi')^2 + \frac{g^2}{8}\Sigma_{ij}(\eta_i{}^\dagger\lambda_{ij}\eta_j - \eta_i'^\dagger\lambda_{ij}\eta_j'$$
$$+ \rho_i{}^\dagger\lambda_{ij}\rho_j - \rho_i'^\dagger\lambda_{ij}\rho_j' + \chi_i{}^\dagger\lambda_{ij}\chi_j - \chi_i'^\dagger\lambda_{ij}\chi_j')^2 \tag{20}$$

$$V_{soft} = m_\eta^2\,\eta^\dagger\eta + m_\rho^2\,\rho^\dagger\rho + m_\chi^2\,\chi^\dagger\chi + m_\eta'^2\,\eta'^\dagger\eta' + m_\rho'^2\,\rho'^\dagger\rho' + m_\chi'^2\,\chi'^\dagger\chi'$$
$$+ B\mu_\eta\,\eta\eta' + B\mu_\rho\,\rho\rho' + B\mu_\chi\,\chi\chi' + H.c. \tag{21}$$

The Higgs potential is obtained as

$$V_H = \frac{g_X{}^2}{2}(\rho^\dagger\rho - \rho'^\dagger\rho' - \chi^\dagger\chi + \chi'^\dagger\chi')^2 + \frac{g^2}{8}[\frac{4\Sigma_{\varphi a \varphi b}}{3}\{(\varphi_a{}^\dagger\varphi_a)^2 + (\varphi_a'^\dagger\varphi_a')^2\}$$

$$-\frac{4}{3}\Sigma_{\phi a\phi b}\{(\varphi_a{}^\dagger\varphi_a)(\varphi_b{}^\dagger\varphi_b)+(\varphi_a'^\dagger\varphi_a')(\varphi_b'^\dagger\varphi_b')+(\varphi_a{}^\dagger\varphi_a)(\varphi_a'^\dagger\varphi_a')-(\varphi_a{}^\dagger\varphi_a)(\varphi_b'^\dagger\varphi_b')\}$$

$$+2\Sigma_{\phi a\phi b}\{(\varphi_a{}^\dagger\varphi_b)(\varphi_b{}^\dagger\varphi_a)+(\varphi_a{}^\dagger\varphi_b')(\varphi_b{}^\dagger\varphi_a')+(\varphi_a'^\dagger\varphi_b')(\varphi_b'^\dagger\varphi_a')\}]$$

$$+\Sigma_{\phi a}\{(m_{\phi a}^2+\mu_{\phi a}^2/4)\phi_a^\dagger\phi_a+(m_{\phi'a}^2+\mu_{\phi a}^2/4)\phi_a'^\dagger\phi_a'+B\mu_{\phi a}\varphi_a\varphi_a'+H.c)\} \quad (22)$$

where $\varphi_{a,b}$ and $\varphi'_{a,b}$ ($a\neq b$) denote ($\eta,\rho,\chi;\ \eta',\rho',\chi'$) fields. The mass spectrum of scalar and pseudoscalar Higgs bosons are obtained by making the shift in neutral scalar fields as

$X_i^0 = \frac{1}{\sqrt{2}}(v_{Xi}+\xi_{Xi}+i\zeta_{Xi})$ where $X_i$ corresponds to $\eta,\rho,\chi,\eta',\rho',\chi'$ neutral scalars.

The scalar potential is minimized to zero with respect to vacuum expectation values (VEV) of all scalars. These include $\langle X^0\rangle = v_X = v,u,V,v',u'$ and $V'$ for $X = \eta,\rho,\chi,\eta',\rho',\chi'$. The gauge boson masses of the SUSY 3-3-1 model are [9]

$$M^2_{W^\pm} = g^2/4\ (v^2+u^2+v'^2+u'^2);\quad M^2_{Y^\pm} = g^2/4\ (v^2+V^2+v'^2+V'^2) \quad (23)$$

$$M^2_{Y^{\pm\pm}} = g^2/4\ (V^2+u^2+V'^2+u'^2);\ M_Z^2 = M_W^2/\cos^2\theta_W \quad (24)$$

$$M_Z^2 = M_W^2/\cos^2\theta_W;\ M_{Z'}^2 = g^2/3\left\{\frac{c_W^2}{(1-4s_W^2)}(V^2+V'^2)+\frac{(1+2s_W^2)^2}{4c_W^2(1-4s_W^2)}(u^2+u'^2)+\frac{(1-4s_W^2)}{4c_W^2}(v^2+v'^2)\right\} \quad (25)$$

The minimization of the effective scalar potential $\partial V_H/\partial v_X = 0$ gives the constraint equations.

$$g^2/12\ (2v^2-u^2-V^2-v'^2+u'^2+V'^2)+m_\eta^2+\mu_\eta^2/4+1/2\ B\mu_\eta\ v'/v = 0 \quad (26)$$

$$g^2/12\ (2u^2-v^2-V^2-u'^2+v'^2+V'^2)+g_X^2(u^2-u'^2-V^2+V'^2)$$
$$+\mu_\rho^2/4+m_\rho^2+1/2\ B\mu_\rho\ u'/u = 0 \quad (27)$$

$$g^2/12(2V^2-v^2-u^2-V'^2+v'^2+u'^2)+g_X^2(V^2-V'^2-u^2+u'^2)+$$
$$\mu_\chi^2/4+m_\chi^2+1/2\ B\mu_\chi\ V'/V = 0 \quad (28)$$

$$g^2/12(2v'^2-u'^2-V'^2-v^2+u^2+V^2)+\mu_\eta^2/4+m_\eta'^2+1/2B\mu_\eta\ v/v' = 0 \quad (29)$$

$$g^2/12(2u'^2-v'^2-V'^2-u^2+v^2+V^2)+g_X^2(u'^2-u^2-V'^2+V^2)$$
$$+\mu_\rho^2/4+m_\rho'^2+1/2B\mu_\rho\ u/u' = 0 \quad (30)$$

$$g^2/12(2V'^2-v'^2-u'^2-V^2+v^2+u^2)+g_X^2(V'^2-V^2+u^2-u'^2)$$
$$+\mu_\chi^2/4+m_\chi'^2+1/2\ B\mu_\chi\ V/V' = 0 \quad (31)$$

These equations constrain the parameters according to

$$\frac{g^2}{12}(v^2 + v'^2) + \mu_\eta^2/2 + m_\eta^2 + m_{\eta'}^2 + \frac{B\mu_\eta}{\sin 2\alpha_1} = 0 \tag{32}$$

$$\frac{g^2}{12}(u^2 + u'^2) + \mu_\rho^2/2 + m_\rho^2 + m_{\rho'}^2 + \frac{B\mu_\rho}{\sin 2\alpha_2} = 0 \tag{33}$$

$$\frac{g^2}{12}(V^2 + V'^2) + \mu_\chi^2/2 + m_\chi^2 + m_{\chi'}^2 + \frac{B\mu_\chi}{\sin 2\alpha_3} = 0 \tag{34}$$

The ratio of the VEV for two neutral Higgses, in 3-3-1 model corresponds to three parameters,

$$\tan \alpha_1 = v'/v \ ; \ \tan \alpha_2 = u'/u \ ; \ \tan \alpha_3 = V'/V \tag{35}$$

These equations can be obtained in terms of masses of gauge bosons

$$M^2_{W^\pm}/3 + \mu_\eta^2/2 + \mu_\rho^2/2 + m_\eta^2 + m_\eta'^2 + m_\rho^2 + m_\rho'^2 + \frac{B\mu_\eta}{\sin 2\alpha_1} + \frac{B\mu_\rho}{\sin 2\alpha_2} = 0 \tag{36}$$

$$M^2_{Y^\pm}/3 + \mu_\eta^2/2 + \mu_\chi^2/2 + m_\chi^2 + m_\chi'^2 + m_\eta^2 + m_\eta'^2 + \frac{B\mu_\eta}{\sin 2\alpha_1} + \frac{B\mu_\chi}{\sin 2\alpha_3} = 0 \tag{37}$$

$$M^2_{Y^{\pm\pm}}/3 + \mu_\rho^2/2 + \mu_\chi^2/2 + m_\rho^2 + m_\rho'^2 + m_\chi^2 + m_\chi'^2 + \frac{B\mu_\rho}{\sin 2\alpha_2} + \frac{B\mu_\chi}{\sin 2\alpha_3} = 0 \tag{38}$$

### 5.1: Mass spectrum for CP-even and CP-odd neutral scalars

The mass squared matrix can be obtained from the quadratic part of $V_H$ in eqn(22) as

$$V_H = \frac{1}{2} m_{ab}^2 \phi_a \phi_b; \text{ where } m_{ab}^2 = \left\langle \frac{\partial^2 V_H}{\partial \phi_a \partial \phi_b} \right\rangle \tag{39}$$

The basis of CP-even sector is defined as $\phi_a = (\xi_\eta, \xi_\eta'; \xi_\rho, \xi_\rho'; \xi_\chi, \xi_\chi')$. We denote the SUSY breaking bilinear terms $B\mu_\phi = -M_{3\phi}^2$. This gives a 6 x 6 mass squared matrix where

$$M^2 = \begin{pmatrix} M_1^2 & M_{12}^2 & M_{13}^2 \\ M_{21}^2 & M_2^2 & M_{23}^2 \\ M_{31}^2 & M_{32}^2 & M_3^2 \end{pmatrix} \tag{40}$$

$$M_1^2 = \begin{pmatrix} \frac{g^2 v^2}{3} + \frac{1}{2}M_{3\eta}^2 \frac{v'}{v} & -\frac{g^2 v v'}{6} - \frac{1}{2}M_{3\eta}^2 \\ -\frac{g^2 v v'}{6} - \frac{1}{2}M_{3\eta}^2 & \frac{g^2 v'^2}{3} + \frac{1}{2}M_{3\eta}^2 \frac{v}{v'} \end{pmatrix} \quad M_{12}^2 = \begin{pmatrix} -\frac{g^2 u v}{6} & \frac{g^2 v u'}{6} \\ \frac{g^2 u v'}{6} & -\frac{g^2 u' v'}{6} \end{pmatrix}$$

$$M_{21}^2 = \begin{pmatrix} -\frac{g^2 u v}{6} & \frac{g^2 v' u}{6} \\ \frac{g^2 u' v}{6} & -\frac{g^2 v' u'}{6} \end{pmatrix} \quad M_2^2 = \begin{pmatrix} \frac{(g^2 + g_X^2)u^2}{3} + \frac{1}{2}M_{3\rho}^2 \frac{u'}{u} & -\frac{(g^2+g_X^2) u u'}{6} - \frac{1}{2}M_{3\rho}^2 \\ -\frac{(g^2+g_X^2) u u'}{6} - \frac{1}{2}M_{3\rho}^2 & \frac{(g^2 + g_X^2)u'^2}{3} + \frac{1}{2}M_{3\rho}^2 \frac{u}{u'} \end{pmatrix}$$

$$M_{31}^2 = \begin{bmatrix} -\frac{g^2 vV}{6} & \frac{g^2 v'V}{6} \\ -\frac{g^2 vV'}{6} & -\frac{g^2 v'V'}{6} \end{bmatrix} \qquad M_{13}^2 = \begin{bmatrix} -\frac{g^2 vV}{6} & -\frac{g^2 v'V}{6} \\ \frac{g^2 v'V}{6} & -\frac{g^2 v'V'}{6} \end{bmatrix}$$

$$M_{32}^2 = \begin{bmatrix} -\frac{(g^2+g_X^2)uV}{6} & \frac{(g^2+g_X^2)u'V}{6} \\ \frac{(g^2+g_X^2)uV'}{6} & -\frac{(g^2+g_X^2)u'V'}{6} \end{bmatrix} \qquad M_{23}^2 = \begin{bmatrix} -\frac{(g^2+g_X^2)uV}{6} & \frac{(g^2+g_X^2)uV'}{6} \\ \frac{(g^2+g_X^2)u'V}{6} & -\frac{(g^2+g_X^2)u'V'}{6} \end{bmatrix}$$

$$M_3^2 = \begin{bmatrix} \frac{(g^2+g_X^2)V^2}{3} + \frac{1}{2}M_{3\chi}^2\frac{V'}{V} & -\frac{(g^2+g_X^2)VV'}{6} - \frac{1}{2}M_{3\chi}^2 \\ -\frac{(g^2+g_X^2)VV'}{6} - \frac{1}{2}M_{3\chi}^2 & \frac{(g^2+g_X^2)V'^2}{3} + \frac{1}{2}M_{3\chi}^2\frac{V}{V'} \end{bmatrix}$$

- **Mass spectrum for CP-even neutral scalars.**

The CP-even real sector does not have zero mass eigenvalues and the diagonalization is done for 2 x 2 submatrices $M_1^2$, $M_2^2$ and $M_3^2$. The physical mass eigenstates include

$H_\varphi = (\cos\theta_i \xi_\varphi' + \sin\theta_i \xi_\varphi)$ ; $h_\varphi = (-\sin\theta_i \xi_\varphi' + \cos\theta_i \xi_\varphi)$; $\varphi = \eta, \rho, \chi$ and $i = 1,2,3$.

The mixing angles ($\theta_1, \theta_2, \theta_3$) are obtained as

$$\tan 2\theta_1 = \tan 2\alpha_1 \frac{[g^2/6 (v^2 + v'^2) + M^2_{A\eta}]}{M^2_{A\eta} - g^2/3(v^2 + v'^2)} \tag{41}$$

$$\tan 2\theta_2 = \tan 2\alpha_2 \frac{[(g^2/6 + g_X^2)(u^2 + u'^2) + M^2_{A\rho}]}{M^2_{A\rho} - (g^2/3 + g_X^2)(u^2 + u'^2)} \tag{42}$$

$$\tan 2\theta_3 = \tan 2\alpha_3 \frac{[(g^2/6 + g_X^2)(V^2 + V'^2) + M^2_{A\chi}]}{M^2_{A\chi} - (g^2/3 + g_X^2)(V^2 + V'^2)} \tag{43}$$

where $M^2_{A\eta} = M^2_{3\eta}/\sin 2\alpha_1$; $M^2_{A\rho} = M^2_{3\rho}/\sin 2\alpha_2$; $M^2_{A\chi} = M^2_{3\chi}/\sin 2\alpha_3$. $\qquad$ (44)

The scalar Higgses $H_\eta^0$, $h_\eta^0$ have eigenvalues from the mass-squared $M_{11}^2$ matrix

$$m^2(H_\eta^0, h_\eta^0) = 1/2 \left[ \frac{g^2}{3}(v^2+v'^2) + M_{A\eta}^2 \pm \sqrt{\{\frac{g^2}{3}(v^2+v'^2) - M_{A\eta}^2\}^2 \cos^2 2\alpha_1 + \sin^2 2\alpha_1\{\frac{g^2}{6}(v^2+v'^2) + M_{A\eta}^2\}^2} \right] \tag{45}$$

The scalars $H_\rho^0$, $h_\rho^0$ are with eigenvalues of mass-squared matrix $M_{22}^2$

$$m^2(H_\rho^0, h_\rho^0) = 1/2 \left[ (g_-^2 + g_X^2)(u^2+u'^2)/3 + M_{A\rho}^2 \pm \right. \qquad (46)$$

$$\left. \sqrt{\{(g_-^2+g_X^2)(u^2+u'^2)/3 - M_{A\rho}^2\}^2 \cos^2 2\alpha_2 + \sin^2 2\alpha_2 \{(g^2+g_X^2)(u^2+u'^2)/6 + M_{A\rho}^2\}^2} \right]$$

The eigenvalues of mass-squared matrix $M_{33}^2$ are for the heaviest Higgses $H_\chi^0$, $h_\chi^0$

$$m^2(H_\chi^0, h_\chi^0) = 1/2 \left[ (g_-^2 + g_X^2)(V^2+V'^2)/3 + M_{A\chi}^2 \pm \right. \qquad (47)$$

$$\left. \sqrt{\{(g_-^2+g_X^2)(V^2+V'^2)/3 - M_{A\chi}^2\}^2 \cos^2 2\alpha_3 + \sin^2 2\alpha_3\{(g^2+g_X^2)(V^2+V'^2)/6 + M_{A\chi}^2\}^2} \right]$$

- **Mass spectrum of CP-odd neutral scalars.**

The mass squared matrices of the CP-odd scalars are obtained from $V_H$ with the basis

$$\varphi_a = (\zeta_\eta, \zeta_\eta'; \zeta_\rho, \zeta_\rho'; \zeta_\chi, \zeta_\chi')$$

The mass matrix obtained after imposing the constraint equations is

$$M_A^2 = \begin{pmatrix} M_{11}^2 & 0 & 0 \\ 0 & M_{22}^2 & 0 \\ 0 & 0 & M_{33}^2 \end{pmatrix}$$

$$M_{11}^2 = -B\mu_\eta \begin{pmatrix} \frac{v'}{v} & 1 \\ 1 & \frac{v}{v'} \end{pmatrix} \qquad M_{22}^2 = -B\mu_\rho \begin{pmatrix} \frac{u'}{u} & 1 \\ 1 & \frac{u}{u'} \end{pmatrix} \qquad M_{33}^2 = -B\mu_\chi \begin{pmatrix} \frac{V'}{V} & 1 \\ 1 & \frac{V}{V'} \end{pmatrix}$$

Diagonalizing the mass matrices, we obtain three Goldstone bosons $G^0_1, G^0_2$ and $G^0_3$

$G^0_1 = -\zeta_\eta \cos\alpha_1 + \zeta_\eta' \sin\alpha_1$ ; $A_\eta = \zeta_\eta \sin\alpha_1 + \zeta_\eta' \cos\alpha_1$, neutral CP = -1 Higgs,

$$M^2_{A\eta} = -B\mu_\eta / \sin 2\alpha_1 = M_{3\eta}^2 / \sin 2\alpha_1 \qquad (48)$$

$G^0_2 = -\zeta_\rho \cos\alpha_2 + \zeta_\rho' \sin\alpha_2$ ; $A_\rho = \zeta_\rho \sin\alpha_2 + \zeta_\rho' \cos\alpha_2$, neutral CP = -1 Higgs,

$$M^2_{A\rho} = -B\mu_\rho / \sin 2\alpha_2 = M_{3\rho}^2 / \sin 2\alpha_2 \qquad (49)$$

$G^0_3 = -\zeta_\chi \cos\alpha_3 + \zeta_\chi' \sin\alpha_3$; $A_\chi = \zeta_\chi \sin\alpha_3 + \zeta_\chi' \cos\alpha_3$, neutral CP = -1 Higgs,

$$M^2_{A\chi} = -B\mu_\chi / \sin 2\alpha_3 = M_{3\chi}^2 / \sin 2\alpha_3 \qquad (50)$$

For gauge boson masses in eqn[36-38]

$$M^2_{W^\pm}/3 = -(\mu_\eta^2/2 + \mu_\rho^2/2 + m_\eta^2 + m_\eta^{'2} + m_\rho^2 + m_\rho^{'2}) + M^2_{A\eta} + M^2_{A\chi}$$

Similar relations exist for $M^2_{Y^\pm}$ and $M^2_{Y^{\pm\pm}}$

### 5.2: Mass spectrum for charged scalars

The 3-3-1 model includes several charged scalars which can be classified as

- charge Q = +2, -2 scalars in the basis ($\rho^{++}, \rho^{'--*}, \chi^{--*}, \chi^{'++}$).

- Charge Q = +1, -1 scalars in the basis ($\rho^+, \rho^{'-*}, \eta_1^+, \eta_1^{'-*}$) ; ($\chi^-, \chi^{'+*}, \eta_2^{+*}, \eta_2^{'-}$)

The 4 x 4 mass-squared matrix in the Q = +2, -2 case is given as

$$M^{++2} = \begin{pmatrix} (\frac{g^2 V^2}{8} + \frac{1}{2} M^2_{3\rho}\frac{u'}{u}) & -\frac{M^2_{3\rho}}{2} & \frac{g^2(uV + u'V')}{8} & 0 \\ -\frac{M^2_{3\rho}}{2} & (\frac{g^2 V^{'2}}{8} + \frac{1}{2} M^2_{3\rho}\frac{u}{u'}) & 0 & \frac{g^2(uV + u'V')}{8} \\ \frac{g^2(uV + u'V')}{8} & 0 & (\frac{g^2 u^2}{8} + \frac{1}{2} M^2_{3\chi}\frac{V'}{V}) & -\frac{M^2_{3\chi}}{2} \\ 0 & \frac{g^2(uV + u'V')}{8} & -\frac{M^2_{3\chi}}{2} & (\frac{g^2 u^{'2}}{8} + \frac{1}{2} M^2_{3\chi}\frac{V}{V'}) \end{pmatrix} \quad (51)$$

The physical mass eigenstates include

$$h_1^{++} = -(\rho^{'--})^* \cos\beta + \rho^{++} \sin\beta \;;\; H_1^{++} = (\rho^{'--})^* \sin\beta + \rho^{++} \cos\beta$$

$$G_2^{--} = -(\chi^{'++})^* \cos\alpha_3 + \chi^{--} \sin\alpha_3 \;;\; H_2^{--} = (\chi^{'++})^* \sin\alpha_3 + \chi^{--} \cos\alpha_3 \quad (52)$$

The diagonalization of 2 x 2 submatrix in upper-left sector yield the mass eigenvalues,

$$m^2_{H1,h1^{++}} = \frac{1}{2}\left[\frac{g^2(V^2 + V^{'2})}{8} + \frac{1}{2} M^2_{3\rho}(\frac{u'}{u} + \frac{u}{u'}) \pm \sqrt{\{(\frac{g^2(V^2 - V^{'2})}{8}) + \frac{M^2_{3\rho}}{2}(\frac{u'}{u} - \frac{u}{u'})\}^2 + M^4_{3\rho}}\right]$$

$$= \frac{1}{2}\left[\frac{g^2(V^2 + V^{'2})}{8} + M^2_{A\rho} \pm \sqrt{\{\frac{g^2(V^2 + V^{'2})\cos 2\alpha_3 - M^2_{A\rho}\cos 2\alpha_2}{8}\}^2 + M^4_{A\rho}\sin^2 2\alpha_2}\right] \quad (53)$$

From the lower right-sector, since $g^2 u^2 << M^2_{3\chi}$, we obtain a Goldstone boson $G_2^{--}$ and doubly charged Higgs $H_2^{--}$, where $m^2_{H_2^{--}} = M_{A\chi}^2$; $m^2_{G_2^{--}} = 0$.

For the singly charged Higgses, $Q = +1$ in the basis $(\rho^+, \rho'^{-*}, \eta_1^{-*}, \eta_1'^+)$, mass-squared matrix is

$$M^{+2} = \begin{pmatrix} (\frac{g^2}{8}v^2 + \frac{1}{2}M^2_{3\rho}\frac{u'}{u}) & -\frac{M^2_{3\rho}}{2} & \frac{g^2}{8}(uv + u'v') & 0 \\ -\frac{M^2_{3\rho}}{2} & (\frac{g^2}{8}v'^2 + \frac{1}{2}M^2_{3\rho}\frac{u}{u'}) & 0 & \frac{g^2}{8}(uv + u'v') \\ \frac{g^2}{8}(uv + u'v') & 0 & (\frac{g^2}{8}u^2 + \frac{1}{2}M^2_{3\eta}\frac{v'}{v}) & -\frac{M^2_{3\eta}}{2} \\ 0 & \frac{g^2}{8}(uv + u'v') & -\frac{M^2_{3\eta}}{2} & (\frac{g^2}{8}u'^2 + \frac{1}{2}M^2_{3\eta}\frac{v}{v'}) \end{pmatrix} \quad (54)$$

The physical mass eigenstates for 2 x 2 upper-left submatrix include

$$h_3^+ = -(\rho'^-)^* \cos\delta + \rho^+ \sin\delta \, ; \, H_3^+ = (\rho'^-)^* \sin\delta + \rho^+ \cos\delta \quad (55)$$

$$m^2_{H_3^+, h_3^+} = \frac{1}{2}\left[\frac{g^2}{8}(v^2 + v'^2) + \frac{1}{2}M_{3\rho}^2(\frac{u'}{u} + \frac{u}{u'}) \pm \sqrt{\{\frac{g^2}{8}(v^2 - v'^2) + \frac{M_{3\rho}^2}{2}(\frac{u'}{u} - \frac{u}{u'})\}^2 + M_{3\rho}^4}\right]$$

$$= \frac{1}{2}\left[\frac{g^2}{8}(v^2 + v'^2) + M_{A\rho}^2 \pm \sqrt{\{\frac{g^2}{8}(v^2 + v'^2)\cos 2\alpha_1 - M_{A\rho}^2 \cos 2\alpha_2\}^2 + M_{A\rho}^4 \sin^2 2\alpha_2}\right]$$

For the lower-right submatrix, the physical mass eigen-states include
$$G_4^+ = -\eta_1'^+ \cos\alpha_1 + (\eta_1^-)^* \sin\alpha_1 \, ; \, H_4^+ = \eta_1'^+ \sin\alpha_1 + (\eta_1^-)^* \cos\alpha_1 \quad (56)$$
For $g^2 u^2 \ll M^2_{3\eta}$, we obtain a Goldstone boson $G_4^+$ and a massive Higgs $H_4^+$, where
$$m^2_{G_4^+} = 0; \, m^2_{H_4^+} = M_{A\eta}^2 \quad (57)$$
For the singly charged scalars ($Q = -1$), in the basis $(\chi^-, \chi'^{+*}; \eta_2^-, \eta_2'^{+*})$,

$$M^{-2} = \begin{pmatrix} (\frac{g^2}{8}v^2 + \frac{1}{2}M^2_{3\chi}\frac{V'}{V}) & -\frac{M^2_{3\chi}}{2} & \frac{g^2}{8}(vV + v'V') & 0 \\ -\frac{M^2_{3\chi}}{2} & (\frac{g^2}{8}v'^2 + \frac{1}{2}M^2_{3\chi}\frac{V}{V'}) & 0 & \frac{g^2}{8}(vV + v'V') \\ \frac{g^2}{8}(vV + v'V') & 0 & (\frac{g^2}{8}V^2 + \frac{1}{2}M^2_{3\eta}\frac{v'}{v}) & -\frac{M^2_{3\eta}}{2} \\ 0 & \frac{g^2}{8}(vV + v'V') & -\frac{M^2_{3\eta}}{2} & (\frac{g^2}{8}V'^2 + \frac{1}{2}M^2_{3\eta}\frac{v}{v'}) \end{pmatrix}$$

The physical mass eigenstates obtained for the upper –left submatrix include

$$G_5^- = -(\chi'^+)^* \cos\alpha_3 + \chi^- \sin\alpha_3 \ ; \ H_5^- = (\chi'^+)^* \sin\alpha_3 + \chi^- \cos\alpha_3 \tag{58}$$

The condition $g^2 v^2 \ll M^2_{3\chi}$ yields a Goldstone boson $G_5^-$ along with a massive $H_5^-$, where $m^2_{G_5^-} = 0; \ m^2_{H_5^-} = M^2_{A\chi}$.

The lower –right submatrix has physical mass-eigenstates

$$h_6^- = -\eta_2'^- \cos\gamma + (\eta_2^+)^* \sin\gamma \ ; \ H_6^- = \eta_2'^- \sin\gamma + (\eta_2^+)^* \cos\gamma \tag{59}$$

$$m^2_{H_6,h_6^-} = \frac{1}{2}\left[\frac{g^2}{8}(V^2+V'^2) + M_{A\eta}^2 \pm \sqrt{\{(\frac{g^2}{8}(V^2+V'^2)\cos 2\alpha_3 - M_{A\eta}^2\cos 2\alpha_1\}^2 + M_{A\eta}^4 \sin^2 2\alpha_1}\right] \tag{60}$$

The mixing angles $\beta, \delta, \gamma$ satisfy

$$\tan 2\beta = \frac{M^2_{A\rho}\tan 2\alpha_2}{M^2_{A\rho} - \frac{g^2 \cos 2\alpha_3 (V^2+V'^2)}{8\cos 2\alpha_2}}$$

$$\tan 2\delta = \frac{M^2_{A\rho}\tan 2\alpha_2}{M^2_{A\rho} - \frac{g^2 \cos 2\alpha_3 (v^2+v'^2)}{8\cos 2\alpha_2}} \quad ; \quad \tan 2\gamma = \frac{M^2_{A\rho}\tan 2\alpha_1}{M^2_{A\rho} - \frac{g^2 \cos 2\alpha_3 (V^2+V'^2)}{8\cos 2\alpha_1}} \tag{61}$$

The essential parameters for the 3-3-1 SUSY model include

(i) $v_0^2 = v^2 + v'^2$ ; $u_0^2 = u^2 + u'^2$ ; $V_0^2 = V^2 + V'^2$

(ii) $\tan\alpha_1, \tan\alpha_2$ and $\tan\alpha_3$ (iii) $M_{A\eta}^2, M_{A\rho}^2$ and $M_{A\chi}^2$.

We choose $M_A^2 = M_{A\eta}^2 = M_{A\rho}^2 = M_{A\chi}^2$; $v_0 = 245.89$ GeV; $u_0 = 7.03$ GeV; $V_0 = 1.054$ TeV

$\tan\alpha_1 = 0.086$; $\tan\alpha_2 = 0.034$; $\tan\alpha_3 = 1$

In Table 1 we list the masses of Higgs bosons for different values of $M_A^2$.

Table1

Masses for CP-even neutral Higgs bosons for different input parameters $M_A^2$

| $M_A^2$ TeV$^2$ | $M_{H\eta}$ GeV | $M_{h\eta}$ GeV | $M_{H\rho}$ GeV | $M_{h\rho}$ GeV | $M_{H\chi}$ TeV | $M_{h\chi}$ TeV |
|---|---|---|---|---|---|---|
| 40 | 197.69 | 97.56 | 199.96 | 9.07 | 1.23 | 0.2 |
| 60 | 241.5 | 101.3 | 244.9 | 9.47 | 1.24 | 0.2 |
| 80 | 278.5 | 104.95 | 282.47 | 9.86 | 1.25 | 0.2 |
| 100 | 311.2 | 108.43 | 316.2 | 10.24 | 1.26 | 0.2 |

Table 2

Masses of charged Higgs bosons for different input parameter $M_A^2 (TeV^2)$. Here the masses $M_{H4}^+ = M_{H5}^- = M_{H2}^{++}$. The Higgs boson masses are in GeV.

| $M_A^2$ | $M_{H3}^+$ | $M_{h3}^+$ | $M_{H6}^+$ | $M_{h6}^+$ | $M_{H1}^{++}$ | $M_{h1}^{++}$ | $M_{H4}^+$ |
|---|---|---|---|---|---|---|---|
| 40 | 200 | 55.6 | 263.5 | 171.6 | 263.5 | 171.6 | 200 |
| 60 | 245 | 55.6 | 299.07 | 171.6 | 299.07 | 171.6 | 245 |
| 80 | 283 | 55.7 | 330.82 | 171.6 | 330.82 | 171.6 | 282.8 |
| 100 | 316.4 | 55.8 | 359.78 | 171.6 | 359.78 | 171.6 | 316.23 |

An interesting relation is obtained between masses of charged Higgs scalars, pseudoscalar Higgses and gauge bosons.

$$M^2_{H1^{++}} + M^2_{h1^{++}} = g^2/8 \, V_0^2 + M^2_{A\rho} = 1/4 \, (M^2_{Y^{++}} + M^2_{Y^+} - M^2_{W^+}) + M^2_{A\rho}$$

$$M^2_{H6^-} + M^2_{h6^-} = g^2/8 \, V_0^2 + M^2_{A\eta} = 1/4 \, (M^2_{Y^{++}} + M^2_{Y^+} - M^2_{W^+}) + M^2_{A\eta} \qquad (62)$$

### 6.1: Higgs-boson couplings to gauge bosons

We consider the interactions of Higgs scalars with gauge bosons which is of much phenomenological interest [11,12] in this section. The gauge invariant kinetic term is

$$L^{kin}_{Higgs} = (\Delta_\mu \varphi)^\dagger (\Delta^\mu \varphi) + (\Delta_\mu \varphi')^\dagger (\Delta_\mu \varphi') \qquad (63)$$

The covariant derivative $\Delta_\mu$ in the kinetic term is defined in 3-3-1 model as [8]

$$\Delta_\mu = \partial_\mu - ig/2 \, W_\mu^a T_a - ie A_\mu Q - \frac{ie (T_{3L} - Qs_W^2)}{s_W c_W} Z_\mu + \frac{ig\sqrt{(1-4s_W^2)}}{c_W} [T_{8L} + \frac{\sqrt{3} \, s_W^2}{(1-4s_W^2)} X] Z'_\mu \qquad (64)$$

The Lagrangian can be decomposed into three terms

$$L_{kin}^{Higgs} = L_{VVH} + L_{HHV} + L_{VVHH} \tag{65}$$

where $V_\mu$ represents the 15-plet gauge bosons of $SU(3)_L$. For neutral scalars $(h_i^0, H_i^0)$, $i = \eta, \rho, \chi$.

(1) $L_{VVH\eta} = [W^+_\mu W^{-\nu} + Y^+_\mu Y^{-\nu} + \frac{1}{2c_W^2}\{Z_\mu Z^\nu + \frac{\sqrt{(1-4s_W^2)}}{\sqrt{3}} Z_\mu Z'^\nu + \frac{(1-4s_W^2)}{3} Z'_\mu Z'^\nu\}] g_{VVH\eta} H_\eta \tag{66}$

(2) $L_{VVH\rho} = [W^+_\mu W^{-\nu} + \frac{1}{2c_W^2} Y^{++}_\mu Y^{--\nu} + \frac{1}{2c_W^2}\{Z_\mu Z^\nu + \frac{(1+2s_W^2)}{\sqrt{3}(1-4s_W^2)} Z_\mu Z'^\nu + \frac{(1+2s_W^2)^2}{3(1-4s_W^2)} Z'_\mu Z'^\nu\}] g_{VVH\rho} H_\rho \tag{67}$

(3) $L_{VVH\chi} = [Y^+_\mu Y^{-\nu} + Y^{++}_\mu Y^{--\nu} + \frac{2s_W^4}{3c_W^2(1-4s_W^2)} Z'_\mu Z'^\nu] g_{VVH\chi} H_\chi \tag{68}$

where

$$g_{VVH\eta} = g^2 v \frac{\sin(\theta_1+\alpha_1)}{2\cos\alpha_1}; \quad g_{VVH\rho} = g^2 u \frac{\sin(\theta_2+\alpha_2)}{2\cos\alpha_2}; \quad g_{VVH\chi} = g^2 V \frac{\sin(\theta_3+\alpha_3)}{2\cos\alpha_3}; \tag{69}$$

- The couplings of two CP-even Higgs boson ($h_i$, $H_i$) to $ZZ, WW, Z'Z'$ satisfy

$$g_{VVHi}/g_{VVhi} = \tan(\theta_m + \alpha_m), \, m = 1,2 \text{ for } i = \eta, \rho; \, V = W, Z, Z'. \tag{70}$$

$$g^2_{WWHi} + g^2_{WWhi} = 4c_W^4 (g^2_{ZZHi} + g^2_{ZZhi}); \, i = \eta, \rho \tag{71}$$

- The pseudoscalar (CP-odd) gauge bosons $A_\varphi$, $\varphi = \eta, \rho$ have tree level couplings to $Z, h_\varphi(H_\varphi)$

$g_{H\varphi A\varphi Z}/g_{h\varphi A\varphi Z} = \tan(\theta_m - \alpha_m), \, m = 1,2;$

$g^2_{H\varphi A\varphi Z} + g^2_{h\varphi A\varphi Z} = \frac{g^2}{4\cos^2\theta_W}; \quad g^2_{H\eta A\eta Z'} + g^2_{h\eta A\eta Z'} = \frac{(1-4s_W^2) g^2}{12\cos^2\theta_W}$

$g^2_{H\rho A\rho Z'} + g^2_{h\rho A\rho Z'} = \frac{g^2(1+2s_W^2)^2}{12c_W^2(1-4s_W^2)}; \quad g^2_{H\chi A\chi Z'} + g^2_{h\chi A\chi Z'} = \frac{g^2 c_W^2}{3(1-4s_W^2)}$

$$2\Sigma_\varphi (g^2_{H\varphi ZZ} + g^2_{h\varphi ZZ}) + M_Z^2 \Sigma_\varphi (g^2_{H\varphi A\varphi Z} + g^2_{h\varphi A\varphi Z}) = \frac{g^2 M_Z^2}{\cos^2\theta_W} \tag{72}$$

- The couplings of charged Higgses to neutral gauge boson $V$ $(Z,A)_\mu$

$L_{HHV} = \frac{g}{2c_W} \{\Sigma_{kj} g_{H_k H_j V} V^\mu (H_k i\partial_\mu H_j) + \Sigma_{kj} g_{h_k h_j V} V^\mu (h_k i\partial_\mu h_j) + \Sigma_{kj} g_{h_k H_j V} V^\mu (h_k i\partial_\mu H_j)\}$

There are relations between $g_{HHV}$, $g_{HhV}$ and $g_{HhV}$ for charged scalars $H_i^\pm$, $h_i^\pm$ where $i = \eta, \rho$.

$$g^2_{ZH_3^+ H_3^-} = g^2_{Zh_3^+ h_3^-} \; ; \; g^2_{ZH_3^- h_3^+} = (\tan^2 2\delta) \, g^2_{ZH_3^+ H_3^-} \; ; \; g^2_{ZH_3^+ H_3^-} + g^2_{ZH_3^- h_3^+} = e^2 \cot^2 2\theta_W$$

$$g^2_{AH_3^- h_3^+} + g^2_{AH_3^+ H_3^-} = e^2 \tag{73}$$

The couplings of charged Higgs bosons $g_{HHW}$, $g_{HhW}$ satisfy

$$g^2_{H\rho H_3^+ W^-} + g^2_{h\rho H_3^+ W^-} = \frac{g^2}{2} = g^2_{h\rho h_3^+ W^-} + g^2_{H\rho h_3^+ W^-} \tag{74}$$

- The quartic interaction terms in Lagrangian $L_{VVHH}$ satisfy

$$g^2_{W^+ W^- H_4^+ H_4^-} = \frac{1}{4}g^4 = g^2_{W^+ W^- H\eta H\eta} = g^2_{W^+ W^- h\eta h\eta} = g^2_{W^+ W^- A\eta A\eta}$$

$$g^2_{A\mu A^\nu H_4^+ H_4^-} = g^2_{A\mu A^\nu H_6^+ H_6^-} = g^2_{A\mu A^\nu h_6^+ h_6^-} = g^2/s_W^2$$

$$g^2_{Z\mu Z^\nu H_4^+ H_4^-} = \frac{g^2 c_{2W}^2}{4c_W^2} ; \; g^2_{Z\mu Z^\nu H_6^+ H_6^-} = g^2_{Z\mu Z^\nu h_6^+ h_6^-} = \frac{g^2 s_W^4}{c_W^2} \tag{75}$$

### 6.2: Trilinear Higgs boson self-interactions

We consider trilinear Higgs self-couplings for $(H_\rho, h_\rho)$, $(H_\eta, h_\eta)$ neutral scalar bosons. The self-couplings of the Higgs bosons follow from the scalar potential $V_H$ in eqn.(22) on using physical Higgs fields. We list the coefficients for these couplings below.

$$L_{3H} = g_{HHH} HHH + g_{HHh} HHh + g_{Hhh} Hhh + g_{hhh} hhh \tag{76}$$

where $H, h = H_\rho, h_\rho$ ; $H_\eta, h_\eta$.

$$g_{H_\rho H_\rho H_\rho} = \frac{g^2 u \, s_W^2 \cos 2\theta_2 \sin(\alpha_2 - \theta_2)}{2(1-4s_W^2)} + \frac{g^2 u}{12}[\sin\theta_2 (2\sin^2\theta_2 - \cos^2\theta_2) + \tan\alpha_2 \cos\theta_2 (2\cos^2\theta_2 - \sin^2\theta_2)]$$

$$g_{h_\rho h_\rho h_\rho} = \frac{g^2 u \, s_W^2 \cos 2\theta_2 \cos(\alpha_2 - \theta_2)}{2(1-4s_W^2)} + \frac{g^2 u}{12}[\cos\theta_2 (2\cos^2\theta_2 - \sin^2\theta_2) - \tan\alpha_2 \sin\theta_2 (2\sin^2\theta_2 - \cos^2\theta_2)]$$

$$g_{H_\rho H_\rho h_\rho} = \frac{g^2 u s_W^2}{2(1-4s_W^2)}[-\sin 2\theta_2 \sin(\alpha_2-\theta_2) - \cos 2\theta_2 \cos(\alpha_2-\theta_2)] + \frac{g^2 u}{12}[\cos\theta_2(2\sin^2\theta_2-\cos^2\theta_2)$$
$$-\tan\alpha_2 \sin\theta_2(2\cos^2\theta_2 - \sin^2\theta_2) + 3/2 \sin 2\theta_2 \sin(\alpha_2-\theta_2)]$$

$$g_{H_\rho h_\rho h_\rho} = \frac{g^2 u s_W^2}{2(1-4s_W^2)}[\sin 2\theta_2 \cos(\alpha_2-\theta_2) - \cos 2\theta_2 \sin(\alpha_2-\theta_2)] + \frac{g^2 u}{12}[\cos\theta_2(2\sin^2\theta_2-\cos^2\theta_2)$$
$$-\tan\alpha_2 \sin\theta_2(2\cos^2\theta_2 - \sin^2\theta_2) + 3/2 \sin 2\theta_2 \cos(\alpha_2+\theta_2)]$$

$$g_{H_\eta H_\eta H_\eta} = \frac{g^2 v}{12}[\sin\theta_1 (2\sin^2\theta_1 - \cos^2\theta_1) + \tan\alpha_1 \cos\theta_1 (2\cos^2\theta_1 - \sin^2\theta_1)]$$

$$g_{h_\eta h_\eta h_\eta} = \frac{g^2 v}{12}[\cos\theta_1 (2\cos^2\theta_1 - \sin^2\theta_1) - \tan\alpha_1 \sin\theta_1 (2\sin^2\theta_1 - \cos^2\theta_1)]$$

$$g_{H\eta H\eta h\eta} = \frac{g^2 v}{12}[\cos\theta_1 (2\sin^2\theta_1 - \cos^2\theta_1) - \tan\alpha_1 \sin\theta_1(2\cos^2\theta_1 - \sin^2\theta_1) + 3/2 \sin2\theta_1 \cos(\alpha_1+\theta_1)]$$

$$g_{H\eta h\eta h\eta} = \frac{g^2 v}{12}[\sin\theta_1 (2\cos^2\theta_1 - \sin^2\theta_1) - \tan\alpha_1 \cos\theta_1(2\sin^2\theta_1 - \cos^2\theta_1) + 3/2 \sin2\theta_1 \cos(\alpha_1-\theta_1)] \quad (77)$$

### 6.2 Yukawa Interactions for three-generations of fermions

The 3-3-1 model allows three generations of quarks and leptons to be considered for Higgs- fermion- antifermion couplings unlike MSSM and thus has tree-level Flavour-changing neutral currents. The gauge-invariant Yukawa Interactions are given for quarks as

$$-L_{Yukawa} = \Sigma_\alpha Q_{\alpha L}\{\Sigma_i(k^d_{\alpha i}\eta d_{Ri}^c + k^u_{\alpha i}\rho u_{Ri}^c) + \Sigma_\beta k^D_{\alpha\beta}\chi D_{R\beta}^c\}$$

$$+ Q_{3L}\{\Sigma_i(k^b_i \rho' d_R^c + k^t_i \eta' u_R^c) + k^T \chi' T_R^c\} \quad (78)$$

$$= \Sigma_\alpha \Sigma_i [k^d_{\alpha i}\{\bar{d_i} P_L(d_\alpha \eta^0 + u_\alpha \eta_1^- + D_\alpha \eta_2^+)\} + k^u_{\alpha i}\{\bar{u_i} P_L(d_\alpha \rho^+ + u_\alpha \rho^0 + D_\alpha \rho^{++})\}$$

$$+ \Sigma_\alpha \Sigma_\beta k^D_{\alpha\beta}\bar{D_\beta} P_L(d_\alpha \chi^- + u_\alpha \chi^{--} + D_\alpha \chi^0) + \Sigma_i k_i^b \bar{d_i} P_L(t\rho'^- + b\rho'^0 + T\rho'^{--})$$

$$+ \Sigma_i k_i^t \bar{u_i} P_L(t\eta'^0 + b\eta_1'^+ + T\eta_2'^-) + k^T \bar{T} P_L(t\chi'^+ + b\chi'^{++} + T\chi'^0) + h.c. \quad (79)$$

Here $i = 1,2,3$; $\alpha,\beta = 1,2$ generation indices and $P_{L,R} = \frac{1}{2}(1 \pm \gamma_5)$ is the right and left-handed projection operator. The tree-level couplings $k^q$ are related to quark masses $m_q$ as

$$k_{11}^d = \sqrt{2} m_d/v;\ k_{11}^u = \sqrt{2} m_u/u;\ k_{11}^D = \sqrt{2} m_D/V;\ k_{\alpha i}^d = \sqrt{2} m_d/v;$$

$$k_3^b = \sqrt{2} m_b/u \tan\alpha_2;\ k_3^t = \sqrt{2} m_t/v \tan\alpha_1;\ k^T = \sqrt{2} m_T/V \tan\alpha_3. \quad (80)$$

Here we do not consider mixing between the first two generations.

The physical Higges now have neutral and charged couplings as

$$-L_Y = L_{neutral} + L_{charged}$$

- $L_{neutral} = \frac{\sqrt{2} m_d}{v}\{(\bar{d},\bar{s},\bar{b})P_L(d,s)(\sin\theta_1 H_\eta + \cos\theta_1 h_\eta + i \sin\alpha_1 G_1^0 + i \cos\alpha_1 A_\eta)\}$

$$+ \frac{\sqrt{2} m_u}{u}\{(\bar{u},\bar{c},\bar{t})P_L(u,c)(\sin\theta_2 H_\rho + \cos\theta_2 h_\rho + i \sin\alpha_2 G_2^0 + i \cos\alpha_2 A_\rho)\}$$

$$+ \frac{\sqrt{2} m_D}{V}\{(\bar{D_1},\bar{D_2})P_L(D_1,D_2)(\sin\theta_3 H_\chi + \cos\theta_3 h_\chi + i \sin\alpha_3 G_3^0 + i \cos\alpha_3 A_\chi)\}$$

$$+ \frac{\sqrt{2} m_b}{u \tan\alpha_2}\{(\bar{d},\bar{s},\bar{b})P_L b (\cos\theta_2 H_\rho - \sin\theta_2 h_\rho - i \cos\alpha_2 G_2^0 + i \sin\alpha_2 A_\rho)\}$$

$$+ \frac{\sqrt{2}\, m_t}{v \tan \alpha_1} \{(\bar{u}, \bar{c}, \bar{t}) P_L t (\cos\theta_1 H_\eta - \sin\theta_1 h_\eta - i \cos\alpha_1 G_1^0 + i \sin\alpha_1 A_\eta)\}$$

$$+ \frac{\sqrt{2}\, m_T}{V \tan\alpha_3} \{\bar{T} P_L T (\cos\theta_3 H_\chi - \sin\theta_3 h_\chi - i \cos\alpha_3 G_3^0 + i \sin\alpha_3 A_\chi)\} + h.c \quad (81)$$

$$L^{charged} = \frac{\sqrt{2}\, m_d}{v} \{(\bar{d}, \bar{s}, \bar{b}) P_L (u,c) \eta_1^- + (\bar{d}, \bar{s}, \bar{b}) P_L (D_1, D_2) \eta_2^+\}$$

$$+ \frac{\sqrt{2} m_u}{u} \{(\bar{u}, \bar{c}, \bar{t}) P_L (d, s) \rho^+ + (\bar{u}, \bar{c}, \bar{t}) P_L (D_1, D_2) \rho^{++}\}$$

$$+ \frac{\sqrt{2}\, m_D}{V} \{(\bar{D}_1, \bar{D}_2) P_L (d, s) \chi^- + (\bar{D}_1, \bar{D}_2) P_L (u,c) \chi^{--}\}$$

$$+ \frac{\sqrt{2} m_b}{u \tan\alpha} \{(\bar{d}, \bar{s}, \bar{b}) P_L t \rho'^- + (\bar{d}, \bar{s}, \bar{b}) P_L T \rho'^{--}\}$$

$$+ \frac{\sqrt{2} m_t}{v \tan\alpha_1} \{(\bar{u}, \bar{c}, \bar{t}) P_L b \eta_1'^+ + (\bar{u}, \bar{c}, \bar{t}) P_L T \eta_2'^-\}$$

$$+ \frac{\sqrt{2} m_T}{V \tan\alpha_3} \{\bar{T} P_L t \chi'^+ + \bar{T} P_L b \chi'^{++}\} + h.c. \quad (82)$$

Using the physical Higgs fields the charged current Lagrangian is

$$L^{charged} = \frac{\sqrt{2}\, m_d}{v} \{(\bar{d}, \bar{s}, \bar{b}) P_L (u, c)(\sin\alpha_1 G_4^- + \cos\alpha_1 H_4^-)$$
$$+ (\bar{d}, \bar{s}, \bar{b}) P_L (D_1, D_2)(\sin\gamma h_6^+ + \cos\gamma H_6^+)\}$$

$$+ \frac{\sqrt{2}\, m_u}{u} \{(\bar{u}, \bar{c}, \bar{t}) P_L (d, s)(\sin\delta h_3^+ + \cos\delta H_3^+)$$
$$+ (\bar{u}, \bar{c}, \bar{t}) P_L (D_1, D_2)(\sin\beta h_1^{++} + \cos\beta H_1^{++})\}$$

$$+ \frac{\sqrt{2} m_D}{V} \{(\bar{D}_1, \bar{D}_2) P_L (d, s)(\sin\alpha_3 G_5^- + \cos\alpha_3 H_5^-)$$
$$+ (\bar{D}_1, \bar{D}_2) P_L (u,c)(\sin\alpha_3 G_2^{--} + \cos\alpha_3 H_2^{--})\}$$

$$+ \frac{\sqrt{2}\, m_b}{u \tan\alpha_2} \{(\bar{d}, \bar{s}, \bar{b}) P_L t (-\cos\delta h_3^- + \sin\delta H_3^-)$$
$$+ (\bar{d}, \bar{s}, \bar{b}) P_L T (-\cos\beta h_1^{--} + \sin\beta H_1^{--})\}$$

$$+ \frac{\sqrt{2} m_t}{v \tan\alpha_1} \{(\bar{u}, \bar{c}, \bar{t}) P_L b (-\cos\alpha_1 G_4^+ + \sin\alpha_1 H_4^+)$$
$$+ (\bar{u}, \bar{c}, \bar{t}) P_L T (-\cos\gamma h_6^- + \sin\gamma H_6^-)\}$$

$$+ \frac{\sqrt{2}\, m_T}{V \tan\alpha_3} \{\bar{T} P_L t (-\cos\alpha_3 G_5^+ + \sin\alpha_3 H_5^+)$$
$$+ \bar{T} P_L b (-\cos\alpha_3 G_2^{++} + \sin\alpha_3 H_2^{++})\} + h.c.$$
$$(83)$$

- The Flavour-changing currents at tree level in 3-3-1 model include

$$L_{FCC} = \frac{\sqrt{2} \cos\alpha_1}{v} \{H_4^- \bar{b}(m_d P_L + m_t P_R)(u,c)$$

$$+ \frac{\sqrt{2}}{u}\{ h_3^+ \bar{t}(m_u \sin\delta P_L - \frac{m_b}{\tan\alpha_2} \cos\delta P_R)(d,s)\}$$

$$+ \frac{\sqrt{2}}{u}\{ H_3^+ \bar{t}(m_u \cos\delta P_L + \frac{m_b}{\tan\alpha_2} \sin\delta P_R)(d,s)\} + h.c. \tag{84}$$

- The Higgs-lepton-lepton interactions are

$$-L_Y^{lepton} = \Sigma_a L_{aL}\{\Sigma_b(k_a^e \rho' e_{Ra}^c + k_a^P \chi' P_{Ra}^c)\} \tag{85}$$

where a,b = 1,2,3 are generation matrices.

The charged lepton masses $m_{e,\mu,\tau}$, $m_{E,M,P}$ arise due to VEV of Higgs fields

$$m_e = k_1^e \frac{u'}{\sqrt{2}} = k_1^e \frac{u}{\sqrt{2}}\tan\alpha_2 \; ; \; m_E = k_1^P \frac{V'}{\sqrt{2}} = k_1^P \frac{V}{\sqrt{2}}\tan\alpha_3 \; ;$$

$$m_\mu = k_2^e \frac{u}{\sqrt{2}}\tan\alpha_2 \; ; \; m_M = k_2^P \frac{V}{\sqrt{2}}\tan\alpha_3 \; ;$$

$$m_\tau = k_3^e \frac{u}{\sqrt{2}}\tan\alpha_2 \; ; \; m_P = k_3^P \frac{V}{\sqrt{2}}\tan\alpha_3 \; ;$$

The Yukawa couplings for leptons include

$$- L_Y^l = \frac{\sqrt{2}}{u\tan\alpha_2}(m_e \bar{e} P_L e + m_\mu \bar{\mu} P_L \mu + m_\tau \bar{\tau} P_L \tau)(\cos\theta_2 H_\rho - \sin\theta_2 h_\rho)$$

$$+ \frac{\sqrt{2}}{V\tan\alpha_3}(m_E \bar{E} P_L E + m_M \bar{M} P_L M + m_P \bar{P} P_L P)(\cos\theta_3 H_\chi - \sin\theta_3 h_\chi) \tag{86}$$

### 6.3 Higgs decays H→W$^+$W$^-$, H→$f_i \bar{f_j}$

The neutral CP-even scalars in 3-3-1 model include ($h_\eta, H_\eta$), ($h_\rho, H_\rho$) with decay channels to vector bosons, quarks and leptons. The branching ratios have much phenomenological interest in different scenarios [13].

(a) The decay rate for $H_\eta \to W^+ W^-$ is given as [2]

$$\Gamma_\eta = \frac{g^4 g^2_{H\eta W^+ W^-} m^3_{H\eta}(1 - 4x_W + 12x_W^2)(1 - 4x_W)^{1/2}}{256 \pi m_W^4}$$

where $x_W = m_W^2/m_{H\eta}^2$, $g_{H\eta W^+W^-} = \dfrac{v \sin(\theta_1 + \alpha_1)}{\cos\alpha_1}$ ; for $G_F = g^2/4\sqrt{2} M_W^2$,

$$\Gamma_\eta = \dfrac{G_F^2 \, v^2 \sin^2(\theta_1 + \alpha_1) m^3_{H\eta}( 1- 4x_W +12x_W^2)(1 - 4x_W)^{1/2}}{8\pi\cos^2\alpha_1} \qquad (87)$$

The decay rate for $H_\rho \to W^+W^-$ is

$$\Gamma_\rho = \dfrac{G_F^2 \, u^2 \sin^2(\theta_2 + \alpha_2) m^3_{H\rho}( 1- 4x_W +12x_W^2)(1 - 4x_W)^{1/2}}{8\pi\cos^2\alpha_2} \qquad (88)$$

where $x_W = m_W^2/m_{H\rho}^2$.

(b) The decay rate for $H_\eta \to \bar{d}_i d_\alpha$ where $i=1,2,3$; $\alpha = 1,2$ generation indices is given by

$$\Gamma = \dfrac{N_c g^2_{H\eta d_i d_\alpha} m_{H\eta}}{8\pi}\left\{1 - \dfrac{(m^2_{d\alpha}+m^2_{di})}{m^2_{H\eta}} - \dfrac{2m_{d\alpha}m_{di}}{m^2_{H\eta}}\right\} \sqrt{\left\{1 - \dfrac{2(m^2_{d\alpha}+m^2_{di})}{m^2_{H\eta}} + \dfrac{(m^2_{d\alpha}- m^2_{di})^2}{m^4_{H\eta}}\right\}}$$

where $N_c = 3$. $\hspace{10cm}$ (89)

Table 3

The Higgs scalar couplings $g_{Hf\bar{f}}, g_{hf\bar{f}}$ for $H_i, h_i$ ; $i = \rho, \eta$

| Scalar field $\phi$ | $g_{\phi t \bar{u}\alpha}$ | $g_{\phi b \bar{d}\alpha}$ | $g_{\phi d \bar{d}\alpha}$ | $g_{\phi u \bar{u}\alpha}$ |
|---|---|---|---|---|
| $H_\eta$ | $m_t \cos\theta_1/\sqrt{2}\, v \tan\alpha_1$ | 0 | $m_d \sin\theta_1/\sqrt{2}\, v$ | 0 |
| $h_\eta$ | $-m_t \sin\theta_1/\sqrt{2}\, v \tan\alpha_1$ | 0 | $m_d \cos\theta_1/\sqrt{2}\, v$ | 0 |
| $H_\rho$ | 0 | $m_b \cos\theta_2/\sqrt{2}u \tan\alpha_2$ | 0 | $m_u \sin\theta_2/\sqrt{2}\, u$ |
| $h_\rho$ | 0 | $-m_b\sin\theta_2/\sqrt{2}\, u \tan\alpha_2$ | 0 | $m_u \cos\theta_2/\sqrt{2}\, u$ |

For $H_\rho \to \bar{b} b$,

$$\Gamma_b = \dfrac{N_c g^2_{H\rho bb} m_{H\rho}}{8\pi}\left(1 - \dfrac{4m_b^2}{m^2_{H\rho}}\right)^{3/2}$$

(c) For the leptonic decay mode $H_\rho \to \tau^+\tau^-$

$$\Gamma_e = \frac{g^2_{H\rho\tau\tau} \cos^2\theta_2 \, m_{H\rho}}{8\pi} \left(1 - \frac{4m_\tau^2}{m^2_{H\rho}}\right)^{3/2}$$

.

where $g_{H\rho\tau\tau} = m_\tau/\sqrt{2}\, u \tan\alpha_2$. (90)

The leptonic decays are suppressed relative to heavy quark decay rates for $H_\rho^0$ Higgs boson while $H_\eta$ does not couple to leptons.

### 7. Conclusions

We have developed the supersymmetric version of a 3-3-1 model [8] derived from $SU(4)_{PS} \otimes SU(4)_{L+R}$ and examined the scalar sector of the model. The SUSY versions of the 3-3-1 models[6] require six scalars $\eta,\rho,\chi; \eta',\rho',\chi'$ in $3_L$ and $3_L^*$ representations of $SU(3)_L$ due to anomaly cancellations. In this work, we consider these scalars such that the physical Higgses are obtained by mixing $(\eta,\eta'),(\rho,\rho')$ and $(\chi,\chi')$ in contrast to other SUSY 3-3-1 models. The trilinear Higgs coupling($\eta\rho\chi$) is absent in the present model and the SUSY breaking terms include $B\mu\phi\phi'$. The tree level potential is minimized to obtain six constraint equations for three pairs of Higgses. The mass matrices are derived for light (h) and heavy (H) scalars of six neutral CP-even and CP-odd states as well as the charged states. The masses for Higgs bosons are listed in Tables 1 & 2. The interactions of Higgs scalars to gauge bosons give relations similar to MSSM [11] between couplings $g_{VHH}$, $g_{VVH}$ and $g_{VVHH}$. A special consequence of the 3-3-1 model is that tree-level flavour changing Higgs-quark-antiquark interactions in eqn[80] are obtained from Yukawa couplings for three generations. We present the values for $g_{Hff}$, $g_{hff}$ in Table 3. The decay rates of neutral Higgses are presented at tree level for WW, $b\bar{b}$ and $\tau\bar{\tau}$. The sparticle spectrum for the model will be considered elsewhere [14].

## Acknowledgements

This work was supported by University Grants Commission project No.F5.1.3(168)/2004 (MRP/NRCB).One of us(AD)would like to thank the organizers of the XX SERC School on 'Theoretical High Energy Physics, 2004',I.IT, Kanpur. We thank A. Kundu for useful discussions.

## References

1.H.P.Nilles,Phys.Rep.**110**, 1(1984); H. Haber and G. Kane, Phys.Rep.**117**,75 (1985); J.Rosick,Phys.Rev.**D41**,3464(1990);J.Rosick,hep-ph/9511250; S.Weinberg, *The Quantum Theory of Fields*, (Cambridge University Press, New York,1996).

2. J.F. Gunion, H.E. Haber, G. L. Kane and S. Dawson, *The Higgs Hunters Guide,*(Addition – Wesley,Reading,M.A.(1990) OPAL Collaboration(The LEP Electroweak Woking Group)[arxiv: hep-ex/0412015].The ALEPH,DELPHI,L3 and OPAL Collab.s, Phys.Lett.**B526**,61(2003).

3. F.Pisano and V. Pleitez, Phys.Rev.**D46**, 410(1992); R. Foot, O. F. Hernandez, F. Pisano and V.Pleitez,Phys.Rev.**D47**,4158 (1993);N.T.Anh, N. Anh Ky, H.N.Long ,Int.J.Mod.Phys.**A16**,541 ( 2001);P.Frampton,Phys.Rev.Lett.**69**,2889(1992);

4. W.A.Ponce, J. B. Florez and L. A. Sanchez, Int.J.Mod.Phys.**A17**, 643(2002); W. A. Ponce, Y. Giraldo and L. A. Sanchez, Phys.Rev.**D67**, 075001(2003)..

5 F.Pisano and V.Pleitez, Phys.Rev **D51**, 3865(1995); Fayyazuddin and Riazuddin, hep-ph/0403042; L. A. Sanchez, F. A. Perez and W. A. Ponce,Eur.Phys.J.**C35**,259(2004).

6. J.C.Montero,V.Pleitez and M.C.Rodriguez,Phys.Rev.**D65**,035006(2002);ibid **D70**, 075004(2004) and hep-ph/0406299; T. V. Duong and E. Ma, Phys.Lett.**B316**, 307(1993);

7.V.Pleitez and M.D.Tonasse, Phys. Rev.**D48**, 2353(1993);J.E.Cieza Montalvo and M.D.Tonasse

hep-ph/0008196.

8. S.Sen and A. Dixit, Phys.Rev.**D71**, 035009(2005).

9. M. Capdequi-Peyranere and M. C. Rodriguez, hep-ph/0103013.

10. Alex G. Dias, R. Martinez and V. Pleitez, hep-ph/0407141

11. John.F.Gunion, H. E. Haber, Nucl.Phys.**B272**, 1(1986);for a review see M.Carena and H.E. Haber, Prog.Part.Nucl.Phys.**50**, 63(2003)

12. M.Duhrssen, S.Heinemeyer, H. Logan, D. Rainwater, G .Weiglein and D. Zeppenfield, Phys. Rev.**D70**, 113009(2004).

13. J.L.Diaz-Cruz,R.Noriega-Papaqui and A.Rosado,Phys.Rev.**D69**,095002(2004).

14. S.Sen and A.Dixit (under preparation)